\def\rn{}
\def\nn#1 #2{#2. #1}				
\def\nnn#1 #2 #3{#2. #3. #1}			
\def\nnnn#1 #2 #3 #4{#2. #3. #4 #1}		
\def\nnnnn#1 #2 #3 #4 #5{#2. #3. #4 #5. #1}	
\def\dualand{ and\hbox{ }}				
\def\multiand{, and\hbox{ }}				
\def\rf#1;#2;#3;#4;#5 {{\frenchspacing\par\rn#1, #3 {\bf #4}, #5 (#2). \par}}
\def\rg#1;#2;#3;#4;#5;#6 {{\frenchspacing\par\rn#1, #3 {\bf #4}, #5 (#2). \par}}
\def\rfbook#1;#2;#3;#4;#5 {{\frenchspacing\par\rn#1, {\it #3} (#5, #4, #2).\par}}
\def\rfprep#1;#2;#3 {{\par\frenchspacing\rn#1, #3 (#2).\par}}

\def\preskip {\vskip-0.0cm}
\def\postskip{\vskip+0.1cm}

\def\beq#1{\begin{equation}\label{#1}}
\def\eeq{\end{equation}}
\def\beqa#1{\begin{eqnarray}\label{#1}}
\def\eeqa{\end{eqnarray}}
\def\eq#1{equation~(\ref{#1})}
\def\Eq#1{Equation~(\ref{#1})}

\def\spose#1{\hbox to 0pt{#1\hss}}
\def\simlt{\mathrel{\spose{\lower 3pt\hbox{$\mathchar"218$}} \raise 2.0pt\hbox{$\mathchar"13C$}}}
\def\simgt{\mathrel{\spose{\lower 3pt\hbox{$\mathchar"218$}} \raise 2.0pt\hbox{$\mathchar"13E$}}}
\def\simpropto{\mathrel{\spose{\lower 3pt\hbox{$\mathchar"218$}} \raise 2.0pt\hbox{$\propto$}}}

\def\bt{\begin{tabbing}}
\def\et{\end{tabbing}}
\def\beq#1{\begin{equation}\label{#1}}
\def\eeq{\end{equation}}

\def\sec#1{Section~\ref{#1}}

\def\bfig{\begin{figure}[h] \centerline{\hbox{}}\vfill}
\def\efig{\end{figure}\vfill\newpage}

\def\fig#1{Figure~\ref{#1}}

\def\fig#1{Figure~\ref{#1}}
\def\Fig#1{Figure~\ref{#1}}

\def\dT{\delta {\rm T}}

\def\dl{\Delta  \l}

\def\deg{^{\circ}}

\def\l{\ell}
\def\etal{{\frenchspacing\it et al.}}
\def\ie  {{\frenchspacing\it i.e.}}
\def\eg  {{\frenchspacing\it e.g.}}

\def\dT{\delta {\rm T}}

\def\dl{\Delta  \l}

\documentstyle[prd,aps,epsf,rotate]{revtex}
\begin{document}
\twocolumn[\hsize\textwidth\columnwidth\hsize\csname@twocolumnfalse\endcsname

\title{The Large-Scale Polarization of the Microwave Background and Foreground}

\author{Ang\'elica de Oliveira-Costa$^{1}$, 
       		         Max Tegmark$^{1}$, 
	          Christopher O'Dell$^{2}$,
		       Brian Keating$^{3}$,
		        Peter Timbie$^{4}$,\\
		   George Efstathiou$^{5}$ \&
			George Smoot$^{6}$}
		      
\address{$^{1}$Department of Physics \& Astronomy, University of Pennsylvania, Philadelphia, PA 19104, USA,  angelica@higgs.hep.upenn.edu\\}
\address{$^{2}$Department of Astronomy, University of Massachusetts, Amherst, MA 01003, USA\\}
\address{$^{3}$Department of Physics, California Institute of Technology, Pasadena, CA 91125, USA\\}
\address{$^{4}$Department of Physics, University of Wisconsin, Madison, WI 53706-1390, USA\\}
\address{$^{5}$Institute  of Astronomy, University of Cambridge, Cambridge CB3 OHA, UK\\}
\address{$^{6}$Department of Physics, University of California, Berkeley, CA 94720, USA\\}

\date{\today. Submitted to Phys. Rev. D.}

\maketitle


\begin{abstract}
The DASI discovery of CMB polarization 
has opened a new chapter in cosmology. Most of the useful information 
about inflationary gravitational waves and reionization is on large 
angular scales where Galactic foreground contamination is the worst,
so a key challenge is to model, quantify and remove polarized foregrounds.
We use the POLAR experiment, COBE/DMR and radio surveys to provide the 
strongest limits to date on the $TE$ cross power spectrum of the CMB on 
large angular scales and to quantify the polarized synchrotron radiation, 
which is likely to be the most challenging polarized contaminant for the 
MAP satellite.
We find that the synchrotron $E$- and $B$-contributions are equal to within 10\% 
from $408-820$MHz with a hint of $E$-domination at higher frequencies.
We quantify Faraday Rotation \& Depolarization effects in the two-dimensional
$(\ell,\nu)$-plane and show that they cause the synchrotron polarization 
percentage to drop both towards lower frequencies and towards lower multipoles.
\bigskip
\end{abstract}

]
  

\section{INTRODUCTION}

\begin{figure}[tb]
\preskip
\centerline{\epsfxsize=9.0cm\epsffile{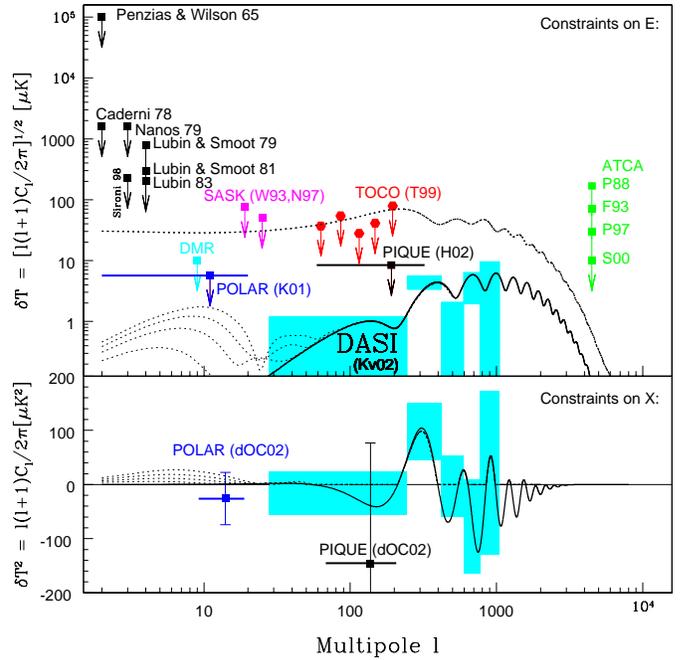}}
\postskip
\caption{\label{summary}\footnotesize%
 	Summary of constraints on polarization so far. 
	From top to bottom, the three curves show the concordance 
	model predictions for $C_\l^T$, $C_\l^E$ and $C_\l^X$, 
	respectively. Four reionization models with $\tau$=0.1, 0.2, 
	0.3 and 0.4 are also plotted (left dotted lines from bottom 
	to top in both plots).
	The limits for $E$ are shown in the upper panel:
	        Penzias \& Wilson 65\protect\cite{PW65},
	        Caderni 78\protect\cite{C78},
	        Nanos   79\protect\cite{N79},
	        Lubin \& Smoot 79\protect\cite{LS79},
	        Lubin \& Smoot 81\protect\cite{LS81},
	        Sironi 98\protect\cite{S98},
	        Lubin  83\protect\cite{L83},
	        SASK (W93\protect\cite{W93},N97\protect\cite{N97}),
	        TOCO (T99 hexagons\protect\cite{T99}),
	        P88\protect\cite{P88},
	        F93\protect\cite{F93},
	        P97\protect\cite{P97},
	        S00\protect\cite{S00},
		DMR\protect\cite{smootDMR},
 	        PIQUE (H02\protect\cite{H02}) and
		POLAR (K01\protect\cite{keating01}).
	The limits for $X$ are shown in the lower panel:
	        PIQUE (d0C02\protect\cite{angel_pique}) and 
	        POLAR (``This Work'').
	The shaded regions are the DASI results 
	(Kv02\protect\cite{kovac02}).
 	}
\end{figure}

The recent discovery of cosmic microwave background (CMB) 
polarization by the DASI experiment \cite{kovac02} 
has opened a new chapter in cosmology -- see \fig{summary}.
Although CMB polarization on degree scales and below can sharpen 
cosmological constraints and provide important cross-checks 
\cite{ZSS97,Eisenstein98}, the potential for the most dramatic improvements 
lies on the largest angular scales where it provides a unique probe of the 
reionization epoch and primordial gravitational waves.
For instance, forecasts \cite{foregpars,knox02} indicate that the MAP satellite 
can measure the reionization optical depth $\tau$ seventeen times more accurately 
using polarization information, and that polarization increases the 
sensitivity of the Planck satellite to tensor modes by a factor of 25. 

Unfortunately, these large scales are also the ones where polarized 
foreground contamination is likely to be most severe, both because of the 
red power spectra of diffuse Galactic synchrotron and dust emission and 
because they require using a large fraction of the sky, including less 
clean patches. The key challenge in the CMB polarization endeavor will 
therefore be modeling, quantifying and removing large-scale polarized 
Galactic foregrounds.
This is the topic of the present paper. We will use the POLAR experiment 
to provide the strongest limits to date on cross-polarized microwave 
background and foreground fluctuations on large angular scales, and employ
polarization sensitive radio surveys to further quantify the polarized 
synchrotron radiation, which is likely to be the most challenging contaminant 
in the polarization maps expected from the MAP satellite.

At microwave frequencies, three physical mechanisms are known to cause 
foreground contamination: synchrotron, free-free and dust emission.
When coming from extragalactic objects, this radiation is usually 
referred to as point source contamination and affects mainly small
angular scales. When coming from the Milky Way, this diffuse Galactic 
emission fluctuates mainly on the large angular scales that are the focus 
of this paper. Except for free-free emission, all the above mechanisms 
are known to emit polarized radiation. 
In the near term, the best measurement of large-scale polarization will 
probably come from the MAP satellite. At MAP's frequency range
(22-90 GHz), synchrotron radiation is likely to be the dominant polarized 
foreground \cite{foregpars}. 
Unfortunately, we still know basically nothing about the polarized 
contribution of the Galactic synchrotron component at CMB frequencies
\cite{foregpars,tucci00,Baccigalupi00,Burigana02,bruscoli02,tucci02,giardino02}, 
since it has only been measured at lower frequencies and extrapolation 
is complicated by Faraday Rotation. This is in stark contrast to the 
CMB itself, where the expected polarized power spectra and their dependence on 
cosmological parameters has been computed from first principles to high 
accuracy \cite{K97,ZS97,Z98,HuWhite97}.
 
Polarization of the Galactic continuum emission was first clearly detected 
in 1962 \cite{Westerhout62}. In the succeeding years, polarization
measurements of the northern sky were made at frequencies between 240 and 
1415~MHz (see \cite{S84} and references therein) with resolutions of only 
a few degrees. No large-area survey has been published since the compendium 
of Brouw and Spoelstra \cite{BS76} and high-resolution surveys have only 
begun to be made recently.
The first major investigation done after \cite{BS76} is that of
\cite{Junkes87}, who observed a section of the Galactic plane defined 
by $49\deg \le \ell \le 76\deg$ and $|b| \le 15\deg$, at frequency of 2.7~GHz. 
The study of \cite{wieringa93} provides the highest resolution insight into 
the small-scale structure of the Galaxy; however, this only covered a few 
areas of the sky which were not larger than a degree or so across. Recently, 
two fully-sampled polarimetric surveys were done at 2.4~GHz 
\cite{D95,D97} and 1.4~GHz \cite{U98,U99}. All of these high-resolution 
surveys covered only regions near the Galactic plane, so in order to use them 
for inferences relevant to CMB experiments, they need to be extrapolated 
both in Galactic latitude and in frequency.
	
The rest of this paper is organized as follows. In section \sec{phenomenology}, 
we review the basics of CMB and synchrotron polarization as well as our 
methods for measuring and modeling it. We present our results in 
\sec{ResultsSec} and discuss our conclusions in \sec{ConclusionsSec}. 


\begin{figure}[tb]
\preskip
\centerline{\epsfxsize=7.5cm\epsffile{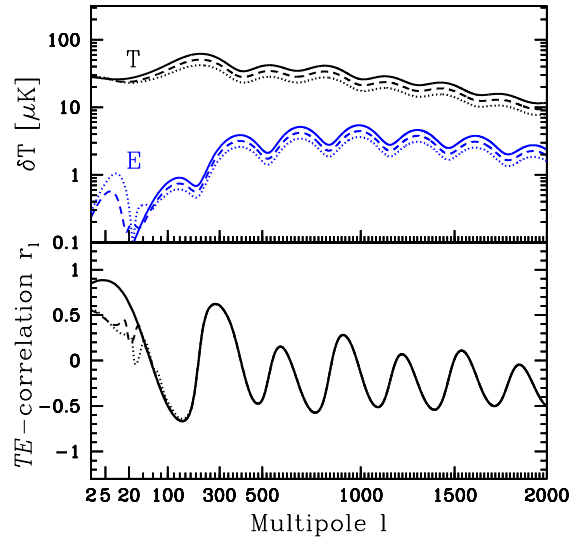}}
\postskip
\postskip
\caption{\label{tauFig}\footnotesize%
	Examples of CMB polarization, showing 
	how the reionization optical depth $\tau$ affects the $T$ and $E$ power 
	spectra (top) and the $TE$ correlation $r_\l$ (bottom). Solid, dashed and 
	dotted curves correspond to $\tau$=0, 0.2 and 0.4, respectively.
	As discussed in \protect\cite{angel_pique}, changing the cosmological 
	parameters affects the polarized and unpolarized power spectra rather 
	similarly except for the cases of reionization and gravitational waves. 
	In the reionization case, a new series of peaks are generated at large 
	scales. 
	$Top$-panel: 
	Although there is no visible change in $T$ at large scales,
        there is clearly a visible change in $E$ since the Sachs-Wolfe nuisance 
	is unpolarized and absent.
	$Lower$-panel:
	On small scales, reionization leaves the correlation $r_\l$ unchanged 
	since $C_\l^T$ and $C_\l^E$ are merely rescaled. On very large scales, 
	$r_\l$ drops since the new polarized signal is uncorrelated with the 
	old unpolarized Sachs-Wolfe signal. On intermediate scales $\l\simgt 20$, 
	oscillatory correlation behavior is revealed for the new peaks.
	For more details about CMB polarization and reionization see 
	\protect\cite{Z97}
        }
\end{figure} 
 
\section{Phenomenology}\label{phenomenology}

\subsection{Notation}\label{notation}

CMB measurements can be decomposed into three maps ($T$,$E$,$B$), 
where $T$ denotes the unpolarized and ($E$,$B$) denote the polarized 
components, respectively. 
Note that an experiment that is insensitive to polarization 
does not measure $T$ but rather that total (unpolarized plus polarized)
intensity; although this distinction has traditionally been neglected 
for CMB experiments where the polarization fraction is small,
it is important both for foregrounds (which can be highly polarized) and
for precision CMB experiments.
From these three maps we can measure a total of 
six angular power spectra, here denoted by 
	$C_\l^T$, 
	$C_\l^E$, 
	$C_\l^B$, 
	$C_\l^X$, 
	$C_\l^Y$ and
	$C_\l^Z$, 
corresponding to the 
	$TT$,  
	$EE$,  
	$BB$,  
	$TE$,  
	$TB$ and  
	$EB$  
correlations\footnote{
	From here on, we adopt the notation
	$TT \equiv T$, 
	$EE \equiv E$,
	$BB \equiv B$, 
	$TE \equiv X$, 
	$TB \equiv Y$ and
	$EB \equiv Z$. 
	},
respectively. 

By parity, $C_\l^Y=C_\l^Z=0$ for scalar CMB fluctuations,  but it is 
nonetheless worthwhile to measure these power spectra as probes 
of both exotic physics  \cite{KK99,XK99,Kamionkowski00} and 
foreground contamination. 
$C_\l^B=0$ for scalar CMB fluctuations to first order in perturbation 
theory \cite{K97,ZS97,Z98,HuWhite97} -- secondary effects such as 
gravitational lensing can create $B$ polarization even if there are 
only density perturbations present  \cite{ZSLENS}. 
In the absence of reionization, $C_\l^E$ is typically a couple 
of orders of magnitude below $C_\l^T$ on small scales and approaches 
zero on the very largest scales. 

The cross-power spectrum $C_\l^X$ is not well suited for the
usual logarithmic power spectrum plot, since it is negative for about 
half of all $\l$-values \cite{angel_pique}. A theoretically more convenient quantity is 
the dimensionless correlation coefficient
\beq{rDefEq}
	r^X_\l\equiv {C_\l^X\over (C_\l^T C_\l^E)^{1/2}},
\eeq
plotted on a linear scale in \fig{tauFig} (lower panel),
since the Schwarz inequality restricts it to lie in the range 
$-1\le r^X_\l\le 1$\footnote{
	Note that for experiments where CMB polarization is measured 
	with a very low signal-to-noise ratio, 
	$C_\l^X$ is a more useful quantity than $r^X_\l$.
	This is because they may be able to place upper and lower
	limits on $C_\l^X$ but can place 
	no meaningful limits on $r^X_\l$ unless they can 
	statistically rule out that $C_\l^E$ in the denominator
	of \eq{rDefEq}.
	}.
From here on we use $r_\l$ as shorthand for $r^X_\l$. For more details
about $r_\l$ and how it depends on cosmological parameters, 
see section II.b in \cite{angel_pique}.


\subsection{Our Knowledge of Synchrotron Emission}\label{synchro}

The Galactic InterStellar Medium (ISM) is a highly complex medium 
with many different constituents interacting through a multitude of 
physical processes. Free electrons spiraling around the Galactic 
magnetic field lines emit synchrotron radiation \cite{rybicki}, 
which can be up to 70\% linearly polarized (see \cite{davies98,smoot99} 
for a review). 

The power spectrum $C_\l$ of synchrotron radiation is normally modeled as a 
power law in both multipole $\l$ and frequency $\nu$, which we will 
parametrize as 
\beq{freq_dep}
    \delta T_\l^2(\nu) = 
    A\left({\l\over 50}\right)^{\beta+2} 
    {\rm with}\quad 
    A\propto \nu^{2\alpha},
\eeq
where $\delta T_\l\equiv [\l(\l+1)C_\l/2\pi]^{1/2}$.
This definition implies that $C_\l\propto\l^\beta$ for $\l\gg 1$
and that the fluctuation amplitude $\propto\nu^{\alpha}$.
The standard assumption is that the total  
intensity has $\alpha\approx -2.8$ with variations of order $0.15$ across the 
sky\footnote{
	Because the spectral index $\alpha$ depends on the energy 
	distribution of relativistic electrons \protect\cite{rybicki}, 
	it may vary somewhat across the sky. One also expects a spectral
	steepening towards higher frequencies, corresponding to a 
	softer electron spectrum (\protect\cite{banday91}; Fig 5.3 in 
	\protect\cite{jonas99}). A recent analysis done at 22~MHz 
	\protect\cite{roger99} shows that $\alpha$ varies slightly over
	a large frequency range.
	}
\cite{platania}.
 
As to the power spectrum slope $\beta$, the 408~MHz Haslam map 
\cite{haslam1,haslam} suggests $\beta$ of order -2.5 to -3.0 down to 
its resolution limit of $\sim 1^\circ$\footnote{
	Although the interpretation is complicated by striping 
	problems \protect\cite{finkbeiner}.
	}
\cite{TE96,bouchet96,bouchet99,newspin}. A similar analysis done on 
the 2.3~GHz Rhodes map of resolution 20$^\prime$ \cite{jonas99} gives 
$\beta = -2.92\pm 0.07$ \cite{giardino01} (flattening to 
$\beta\approx -2.4$ at low Galactic latitudes \cite{giardino02}). 

For the polarized synchrotron component, our observational knowledge is, 
unfortunately, not as complete. 
To date, there are measurements of the polarized synchrotron 
power spectrum obtained basically from three different surveys\cite{reich01}: 
	the Leiden surveys\footnote{
		The observations done by Brouw and Spoelstra covered almost 
		40\% of the sky extending to high Galactic latitudes. Using
		the same instrument, they observed the polarized Galaxy in 
		$Q$ and $U$ in five frequencies from 408~MHz up to 1.4~GHz 
		and with angular resolutions from 2.3$\deg$ at 408~MHz up 
		to 0.6$\deg$ at 1.4GHz. Unfortunately this data was also
		undersampled, making it difficult to draw inferences about 
		its polarized power spectrum.
		} \cite{BS76,S84}, 
	the Parkes 2.4~GHz Survey of the Southern Galactic Plane\footnote{
		This survey covers a strip 127$\deg$ long and at least 
		10$\deg$ wide centered in the Galactic plane, with a resolution 
		of FWHM=10.4$^\prime$. It is publically available at 
     		{\it http://www.uq.edu.au/$\sim$roy/}. 
		} \cite{D95,D97}, and 
	the Medium Galactic Latitude Survey\footnote{
		The Medium Galactic Latitude Survey maps the Galactic plane 
		within $\pm$20$\deg$, with a resolution of FWHM=9.35$^\prime$ 
		at 2.4~GHz. This survey is partially available at 
 		{\it http://www.mpifr-bonn.mpg.de/staff/buyaniker/index.htm}. 
		} \cite{U98,U99,D99}.

These measurements exhibit a much bluer power spectrum in polarization 
than in intensity, with $\beta$ in the range from 1.4 to 1.8 
\cite{foregpars,tucci00,Baccigalupi00,Burigana02,bruscoli02,tucci02,giardino02}. 
These results are usually taken with a grain of salt when it comes to their 
implications for CMB foreground contamination, for three reasons:
\begin{enumerate}
	\item Extrapolations are done from low to high galactic latitudes;
	\item Extrapolations are done from low to high frequencies; and
	\item Much of the available data is undersampled.
\end{enumerate} 
The Leiden surveys extend to high Galactic latitudes and up to 1.4 GHz 
but are unfortunately undersampled, while the Parkes and the Medium 
Galactic Latitude Surveys only probe regions around the Galactic 
plane. In the following three sections, we will discuss these three problems 
in turn.


\subsubsection{The Latitude Extrapolation Problem}

\begin{figure}[tb] 
{\centerline{\epsfxsize=8.5cm\epsffile{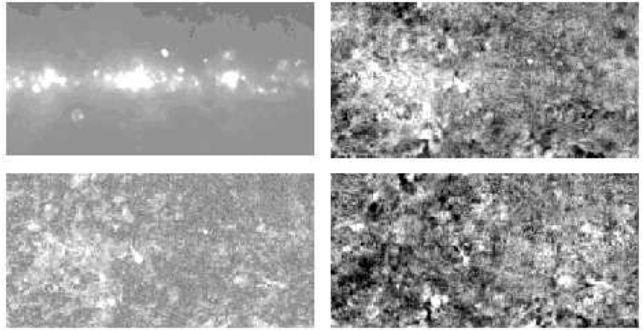}}}
{\vskip+0.3cm} 
\caption{\label{DuncanPIQU}\footnotesize%
	 The nature of the Galactic synchrotron emission. Clockwise from top left, 
	 the panels show Stokes $T$, $U$, $Q$, and $P$ (defined as $P$=$\sqrt{Q^2+U^2}$) 
	 from Block 3 of the Parkes 2.4 GHz Survey of the Southern Galactic 
	 Plane.
         }
\vskip-0.2cm
\end{figure}

\begin{figure}[tb]
\centerline{\epsfxsize=8.5cm\epsffile{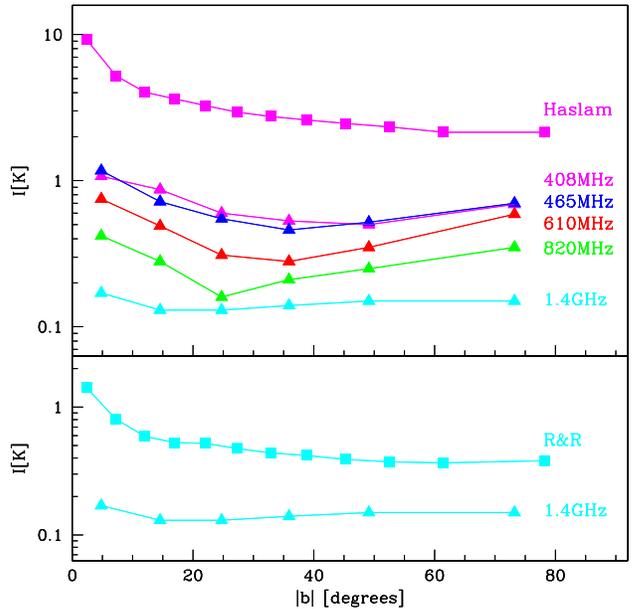}}
\caption{\label{background}\footnotesize%
         The polarized and total synchrotron component as a function
	 of the Galactic latitude.
	 Each of five Leiden polarized surveys was divided in six slices 
	 of equal area, we then calculated the mean intensity (defined as
	 $P$=$\sqrt{Q^2+U^2}$) for each of those slices. A similar procedure
	 was used for the polarization insensitive Haslam and Reich \& Reich surveys, 
	 but 12 slices were chosen instead. 
	 The top panel show the results from the five Leiden surveys plus 
	 the 408~MHz Haslam data, while the bottom panel show the results
	 from the Leiden 1.4GHz survey and 1.42~GHz Reich \& Reich data.
	 Comparison between polarized components and the total intensity 
	 at the same frequency illustrates that 
	 the polarized synchrotron is almost independent 
	 of the Galactic latitude while the unpolarized emission is 
	 strongly concentrated in the Galactic plane.
         }
\end{figure}

Although only high Galactic latitudes are relevant for CMB work, 
most of the data used for understanding the polarized CMB foreground 
contamination is at low Galactic latitudes.
\Fig{DuncanPIQU} shows that whereas the total intensity of the synchrotron 
emission depends strongly on the Galactic latitude, the polarized
component is approximately independent of Galactic latitude --- indeed, in the
three polarized images, it is difficult to distinguish the galactic plane at all.
As noticed long ago by \cite{D97}, there is a faint, quasi-uniform 
polarized component of the Galactic polarized emission in their survey, 
upon which the emission from other features is superimposed: towards 
the higher latitudes, this faint component appears similar in both 
structure and intensity to the correspondent lower latitude emission.
This well-know empirical result can be also seen (in a more quantitative way)
in the Leiden surveys. \Fig{background} shows that in the frequency range between 
408~MHz to 1.4~GHz, the Polarization Intensity $P$ ($P$=$\sqrt{Q^2+U^2}$) is basically 
constant as the Galactic latitude $|b|$ increases, whereas the polarization insensitive surveys 
(such as the 408~MHz Haslam and the 1420~MHz Reich \& Reich \cite{reich88}) have 
the bulk of their emission coming from the Galactic plane.

The usual interpretation of this very weak latitude dependence of polarized
synchrotron radiation is that the signal is dominated by sources that are nearby
compared to the scale height of the Galactic disk, with more distant sources 
being washed out by Depolarization (to which we return in the next subsection).
As a result, having well-sampled polarized maps off the galactic plane at the 
same frequencies would not be expected to affect our results much, since they 
would be similar to those in the plane. This issue, however, deserves more work 
as far as extrapolation to CMB frequencies is concerned: the latitude dependence 
may well return at higher frequencies as Depolarization becomes less important, 
thereby revealing structure from more distant parts of the Galactic plane.
In this case, extrapolating from an observing region around the Galactic plane to 
higher latitudes may well result in less small-scale power in the angular distribution.
 
If we are {\it lucky}, many of the complications of extrapolating to higher latitude
may largely cancel out the complications of extrapolating to higher frequency, 
thereby making it easier to quantify the polarized CMB foreground problem.
The reason for optimism is the following: at high latitudes (which is all that 
really matters for CMB research), the foreground signal will be entirely due to 
nearby emission within the scale height of the thick Galactic disk; and
at low frequencies in the Galactic plane (which is where we have really good data),
the polarized signal we see may well be dominated by such nearby emission,
with emission from more distant regions in the Galactic disk hidden by 
Depolarization.

\smallskip

\setcounter{footnote}{1}


\subsubsection{Faraday Rotation, Depolarization and the Frequency Extrapolation Problem}
\label{faraday}


The plane of a polarized wave may be regarded as the sum of two 
circularly polarized components of opposite handedness. In an ionized 
medium with a non-zero magnetic field, these two 
components propagate with different phase velocities, which will 
result in a rotation of the plane of polarization of the linearly 
polarized radiation. 
This rotation, known as the Faraday Rotation\footnote{
	A detailed discussion of the Faraday Rotation \&
	Depolarization effects as well as their importance
	in astrophysical observations is in given in
	\protect\cite{sokoloff98}.},
produces a change in polarization angle $\Delta \theta$ of 
\beq{phiequation}
    \Delta \theta = 0.81\lambda^2 \int_0^L n_e B_\| dL 
    		  =     \lambda^2 RM~{\rm [rad]}, 
\eeq 
where $\lambda$ is the wavelength given in meters and the quantity 
$\Delta \theta/\lambda^2$ is called the Rotation Measure
($RM$ -- usually expressed in units of rad m$^{-2}$).
The integral is done over the line of sight from us to the 
emitting region at a distance $L$ in pc, $n_e$ is the the 
free electron density in cm$^{-3}$ and $B_\|$ is the magnetic 
field parallel to the line of sight in $\mu$G.

From \eq{phiequation} it is easy to see that observations of 
this synchrotron radiation in several frequencies allows the 
determination of Rotation Measures in the diffuse radiation. 
From the obtained structure in the Rotation Measure on different 
scales, we can obtain information on the magnetic field parallel 
to the line of sight, weighted with electron density --
an example of this method can be found in \cite{tools}.
In radio astronomy, Faraday Rotation has become one of the 
main tools to investigate the interstellar magnetic field
(see, \eg~\cite{Haverkorn00,Gaensler00}). 


It is important to point out, however, that Faraday Rotation can 
only change the polarization angle and not the polarized intensity 
$P$. The fact that we do see structure in $P$ that is not correlated
with a counterpart in intensity $T$ implies that part of the 
radiation has been depolarized \cite{wieringa93}. 
A simple visual comparison of the total intensity and polarized maps 
of the same region in the sky of the Parkes 2.4~GHz survey shows
Depolarization at work (see \Fig{DuncanPIQU}): many sources 
which present an intense total emission do not show a counterpart 
in the polarized maps; similarly bright regions of extended polarization 
are not connected with unpolarized sources. 
A more detailed study of this same survey reached similar
conclusions: Giardino {\etal} \cite{giardino02} showed that the $E$ and 
$B$ power spectra were dominated by changes in the polarization angle 
rather than by changes in the polarized intensity, suggesting that 
Faraday Rotation was playing a significant role\footnote{
	Although at first glance the images in \fig{DuncanPIQU} suggest 
	that the polarized and unpolarized components are uncorrelated, 
	\protect\cite{D97} found that for some patches in their images 
	there is a good correlation between the polarized and total power 
	intensities. Therefore they conclude that a good fraction of the 
	polarized emission seen over the plane was caused by changes in 
	synchrotron emissivity, rather them any Depolarization or Faraday 
	Rotation of the synchrotron background. According to
	\protect\cite{D97}, variations in synchrotron emission can be 
	caused by increases in the density of relativistic electrons (due 
	to SNRs), and/or variations in the magnetic field intensity.
	
	It is important to point out that the relative importance of these 
	two mechanisms (Faraday Rotation \& Depolarization and changes 
	in the synchrotron emissivity of the source regions) over the 
	Galactic plane region are currently unknown \cite{D99b}.
	}.
 
Depending on the frequency and beamwidth used, Depolarization can
play an important role in polarization studies of the Galactic 
radio emission \cite{S84}. As discussed by Cortiglioni and 
Spoelstra \cite{CS95}, Depolarization can have four causes:
  {\it 1)} differential polarization along the line of sight,
  {\it 2)} differential polarization across the beam,
  {\it 3)} differential Faraday Rotation across the beam, and  
  {\it 4)} differential Faraday Rotation and polarization across the bandwidth.  
If the bandwidth is very narrow, we can neglect item {\it 4}; 
also, if the polarized data has been sufficiently sampled, smoothing 
it to a largest beam may inform us about items {\it 2} and {\it 3}, 
leaving us with item {\it 1} as the expected main source of 
Depolarization\footnote{ 
	In the case of Leiden surveys, item {\it 4} is negligible.
	Based in previous analysis done over the Galactic 
	loops at 1.4~GHz \protect\cite{S71,S72}, Spoelstra 
	\protect\cite{S84} argued that items {\it 2} and {\it 3} 
	have a relatively minor contribution to the Depolarization 
	in those surveys. Leaving, therefore, differential 
	polarization along the line of sight as the main source 
	of Depolarization.
	}.

Because of the complicated interplay of these mechanisms, we should expect 
both the amplitude and the shape of the polarized synchrotron power spectrum 
to change with frequency. We will therefore take an empirical approach below
and use the available data to map out (for the first time) the two-dimensional 
region in the $(\ell,\nu)$ plane where Faraday Rotation \& Depolarization 
are important.


\subsubsection{Incomplete Sky Coverage and the Undersampling Problem}

For the case of undersampling in the Leiden surveys, some authors have
overcome this problem by doing their Fourier analysis over selected
patches in the sky where they believe the average grid space in the 
patch is close to the map's beam size, so that they can apply a 
Gaussian smoothing on it -- this is well explained and illustrated 
in \cite{bruscoli02}.
Fortunately, we can eliminate this problem by measuring the power spectra 
with the matrix-based quadratic estimator technique that has recently been 
developed for analyzing CMB maps 
\cite{BJK,TC01,angel_pique}.

Although the undersampling and partial sky coverage results in unavoidable 
mixing between different angular scales $\l$ and polarization types 
($E$ and $B$), this mixing (a.k.a. {\it leakage}) is fully quantified 
by the window functions that our method computes \cite{TC01} 
and can therefore be included in the statistical analysis without 
approximations.
Specifically, we compute the six power spectra ($T,E,B,X,Y,Z$) described 
in \sec{notation} so that the leakage, if any, is minimal.

In \cite{TC01} it was argued that susceptibility to systematic errors 
could be reduced by choosing the ``priors'' that determine the quadratic 
estimator method to have vanishing cross-polarizations, $X=Y=Z=0$, and 
it was shown that this simplification came at the price of a very small 
(percent level) increase in error bars.
In Appendix A of \cite{angel_pique}, it was shown that this choice has an
important added benefit: exploiting a parity symmetry, it eliminates 14 
out of the 15 leakages, with only the much discussed 
	\cite{TC01,Z98,Jaffe00,Zalda01,Lewis02,Bunn01} 
$E-B$ leakage remaining. 
In \cite{Bunn02} it was shown that even the remaining $E-B$ leakage can, 
in principle, be removed. Unfortunately, this technique cannot be 
applied here, since it works only for a fully sampled two-dimensional map.


\bigskip
\bigskip
\bigskip
{\bf \centerline{Table 1 -- POLAR-DMR Power Spectrum}} 
\medskip
\centerline{\label{tab1}
\begin{tabular}{lrrr}
\hline
\hline
\multicolumn{1}{l}{}			          &
\multicolumn{1}{c}{$\ell_{\rm eff}\pm\dl$}	  & 
\multicolumn{1}{c}{$\dT ^2\pm \sigma\> [\mu K^2]$}&
\multicolumn{1}{r}{$\dT\> [\mu K]^{(a)}$}         \\
\hline 
 $T$         &15.6$\pm$6.6     & 487.0$\pm$270.6     &22.1$^{+7.4}_{-5.5}$\\ 
 $E$         &12.6$\pm$4.5     &  -9.9$\pm$ 32.0     &$<$4.7~(7.4)\\ 
 $B$         &12.6$\pm$4.5     &  13.9$\pm$ 32.0     &$<$6.8~(8.8)\\
 $X$         &14.0$\pm$4.8     & -26.0$\pm$ 48.5     &$<$8.7(11.1)\\         
 $Y$         &14.0$\pm$4.8     &  -0.1$\pm$ 48.5     &$<$7.0~(9.8)\\ 	
 $Z$         &11.4$\pm$2.9     & -50.0$\pm$ 31.6     &$<$6.6(10.7)\\
\hline
\hline
\end{tabular} 
} 
\smallskip
 \noindent{\small $^{(a)}$Values in parentheses are 2-$\sigma$ upper limits.
 Cross-correlation upper limits refer to $|X|$, $|Y|$ and $|Z|$.} \\
\bigskip
			
\section{Results}\label{ResultsSec}

\subsection{POLAR Power Spectra}

POLAR was a ground-based CMB polarization experiment that operated
near Madison, Wisconsin \cite{keating01,keating02,odell02}. It used 
a simple drift-scan strategy, with a 7$\deg$ FWHM beam at 26--30~GHz, and  
simultaneously observed the Stokes parameters $Q$ and $U$ in a ring 
of declination $\delta = 43\deg$. Because POLAR was insensitive to 
the unpolarized CMB component, we cross-correlate their $Q$ and $U$ 
data with the $T$-data from the $COBE$/DMR map \cite{smoot92}.

\subsubsection{Quadratic Estimator Analysis}

We measure the six power spectra described in \sec{notation} 
using the quadratic estimator method exactly as described in \cite{TC01}.
We computed fiducial power spectra with the CMBFAST software \cite{SZ96} 
using cosmological parameters from the concordance model from \cite{X01} 
(that of \cite{Efstathiou01} is very similar). Table~1 shows the result 
of our band-power estimation. The values shown in parentheses in the 
rightmost column of this table are our 2-$\sigma$ upper limits.
In these calculations, we used 5 multipole bands of width $\Delta \l = 6$ 
for each of the six polarization types $(T,E,B,X,Y,Z)$, thereby going out 
to $\l = 30$, and average the measurements together with inverse-variance 
weighting into a single number for each polarization type to minimize noise. 

We used the combined DMR 53+90~GHz data to obtain good sensitivity 
to the unpolarized component. 
We perform our analysis using strips of the DMR data of width $\pm 15\deg$ 
around the POLAR declination -- we found that further increasing in the 
width of these disks did not significantly tighten our constraints. 
Finally, we eliminated sensitivity to offsets by projecting out the mean 
(monopole) from the $T$, $Q$ and $U$ maps separately. 
 
The detection of unpolarized power is seen to be consistent with that 
published by the DMR \cite{smoot92} group. Table~1 shows that we detect 
no polarization or cross-polarization of any type, obtaining mere upper 
limits, just as the models predict.
The window functions reveal substantial leakage between $E$ and $B$, so 
that the limits effectively constrain the average of these two spectra 
rather than both separately. This large leakage is due to the one-dimensional 
nature of the POLAR dataset, and can be completely eliminated with a  
fully sampled two-dimensional map \cite{Bunn02}.

Finally, we perform the same analysis described above by replacing the DMR 
stripe with a similar stripe selected from the 408~MHz Haslam map (which 
was smoothed to 7$\deg$ and scaled to 30~GHz using $\beta_T=-$3). We detected 
no cross-polarization of any type between POLAR and the Haslam map, 
obtaining a mere upper limit of 
 $|X| \simlt$11.0$\mu K$ (or a 2-$\sigma$ upper limit of 15.4$\mu K$).


\subsubsection{Likelihood Analysis}

We complement our band-power analysis with a likelihood analysis where 
we assumed that $B=0$. Specifically, we set $B=Y=Z=0$ and take each of 
the remaining power spectra  $(T,E,X)$ to be constant out to $\l=30$. 
 
\begin{figure}[tb]
\centerline{\epsfxsize=9cm\epsfysize=9cm\epsffile{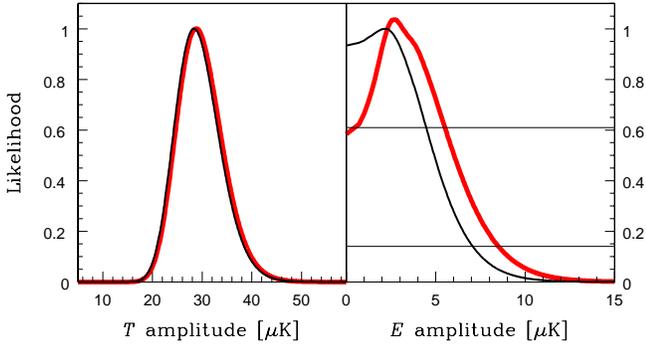}}
{\vskip-4.0cm} 
\caption{\label{limitsTE}\footnotesize%
	 Likelihood results using the $E$-polarized information alone 
	 (right panel, solid black line), using $T$ information alone
	 (left panel, solid black line), and using both POLAR and DMR 
	 $T$-information and marginalizing (solid red lines). From top 
	 to bottom, the two horizontal lines correspond to 68\% and 
	 95\% confidence limits, respectively. 
         }
\end{figure} 

We first perform a simple 1-dimensional likelihood analysis for the parameter 
$E$ using the POLAR data alone (discarding the DMR information), obtaining 
the likelihood function in excellent agreement with that published by 
\cite{keating01} -- see \Fig{limitsTE} (right panel, solid black line). 
A similar 1-dimensional likelihood analysis for the parameter $T$ using 
the DMR data alone produces $T\approx 28\mu$K, consistent with that of
the DMR team \cite{smoot92} (left panel, solid black line).
We then compute the likelihood function including both POLAR and DMR data 
in the 3-dimensional space spanned by $(T,E,r_\ell)$ and compute constraints 
on individual parameters or pairs by marginalizing as in \cite{X01}. 
Once again, we obtain a $T$-measurement in complete agreement with that
for the DMR team (left panel, solid red line).
 
\begin{figure}[tb]
{\vskip-0.2cm}
\centerline{\epsfxsize=9.3cm\epsffile{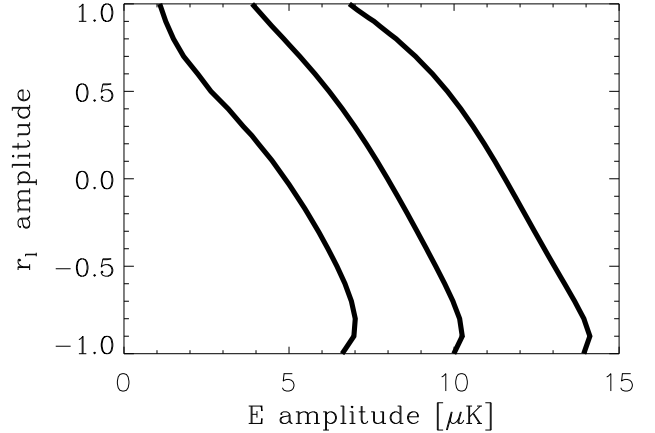}}
\caption{\label{Er}\footnotesize%
	Joint constraints on $E$ polarization and $r_\ell$ after
	marginalizing over $T$. From left to right, the contours show that the 
	likelihood function has dropped to $e^{-1.1}$, $e^{-3.0}$ and $e^{-4.6}$ 
	times its maximum value, which would correspond to 68\%, 95\% and 
	99\% limits if the likelihood were Gaussian. 
	For comparison, the 
	concordance model predicts ($E,r_\ell$)=(0.001,0.66) at $\l$=14, 
	the center of our window function for $X$ (see Table~1).
        }
\end{figure}

\Fig{Er} shows our constraints in the $(E,r_\ell)$-plane after marginalizing 
over $T$. It is seen that our constraints on the cross-polarization 
are weaker than the Schwarz inequality $|r_\ell| \le 1$, so in this sense the 
data has taught us nothing new. The likelihood function is seen to be 
highly non-Gaussian, so obtaining statistically meaningful confidence limits (which
is of course uninteresting in our case, since the constraints are so weak)
would involve numerically integrating the likelihood function.
Since $r_\l$ is expected to oscillate between positive and negative values,
using a flat (constant) $r_\l$ in the likelihood analysis runs the risk of 
failing to detect a signal that is actually present in the data, canceling out 
positive and negative detections at different angular scales. This is not likely 
to have been a problem in our case, since $r_\l$ is uniformly positive in our 
sensitivity range $\l=14\pm 5$ for the concordance model.
 
\Fig{summary} compares our results with all other polarization constraints 
published to date. 

\subsection{The Leiden Power Spectra}  

\subsubsection{Basic power spectra}

For the Leiden surveys, our analysis was performed using 10 multipole bands 
of width $\Delta\l=10$ for each of the six polarization types $(T,E,B,X,Y,Z)$, 
thereby going out to $\l=100$. We used the Haslam map for the unpolarized 
component $T$, scaled and smoothed to match Leiden's five different frequencies. 
A Galactic cut of $|b|=25\deg$ was applied in order to match the POLAR 
observing region. We iterated the QE method once and chose the second 
prior to be a simple power law model consistent with the original measurement
for the $T$, $E$ and $B$ power. The priors for $X$, $Y$ and $Z$ were
set as zero.

\begin{figure}[tb]
\centerline{\epsfxsize=8.5cm\epsffile{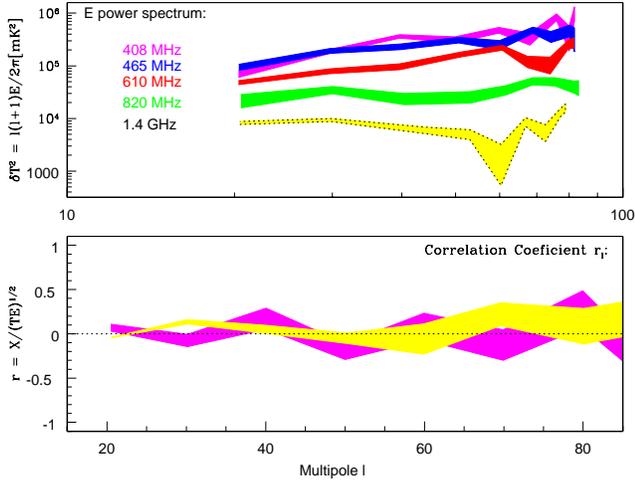}}
{\vskip-1.7cm} 
\caption{\label{power}\footnotesize%
	 Power spectra of the Leiden data. Top panel shows 
	 the $E$ power spectra for the five Leiden frequencies going 
	 from 408 to 1411~MHz, while the Bottom panel shows the $X$ 
	 cross power spectrum $r_\l$ for two of the five Leiden 
	 frequencies (each frequency is represented by the same color 
	 in both plots). Intrinsic $EB$ correlation could be present 
	 but masked by Faraday Rotation, since random rotations 
	 of the polarization angle would cause correlations to average 
	 to zero.
         }
\end{figure}

\Fig{power} shows the $E$ power spectra (top) and $r_\l$ correlation coefficient 
(bottom) of the Leiden surveys. We find that all power spectra are well 
approximated by powers laws as in \eq{freq_dep}. The best fit normalizations 
$A$ and slopes $\beta$ for $E$ and $B$ are shown in Table~2. The values 
of $\beta$ are consistent with previous analyses 
\cite{foregpars,tucci00,Baccigalupi00,Burigana02,bruscoli02,tucci02,giardino02},
showing that the slopes get redder as frequency increases.

\bigskip

{\bf \centerline{Table 2 -- Normalization \& Spectral Index$^{(a)}$}} 
\medskip
\centerline{
\begin{tabular}{ccccc}
\hline
\hline
\multicolumn{1}{c}{$\nu$}       &
\multicolumn{1}{c}{$A_E$}       &
\multicolumn{1}{c}{$\beta_E$}   &
\multicolumn{1}{c}{$A_B$}       &
\multicolumn{1}{c}{$\beta_B$}   \\
 (GHz)    &$[mK^2]$ &	     &$[mK^2]$ &    \\
\hline
 0.408    &5.5      &-0.5    &5.7      &-0.4\\
 0.465    &5.4      &-1.0    &5.4      &-0.5\\
 0.610    &5.1      &-1.0    &5.1      &-0.8\\
 0.820    &4.5      &-1.5    &4.6      &-1.8\\
 1.411    &3.9      &-1.9    &3.6      &-2.6\\
\hline
\hline
\end{tabular} 
}
\medskip
\noindent{\small $^{(a)}$All fits are normalized at $\ell$=50, \ie,
		        $\delta {\rm T}_{\ell}^2 = A (\ell/50)^{\beta+2}$.} 
\medskip
\bigskip

For all Leiden surveys, the $X$ and $Y$ power spectra are 
found to be consistent with zero --
the 2.4~GHz Parkes survey had a similar finding for $X$ \cite{giardino02}.
These are not surprising results: if Faraday Rotation 
makes the polarized and unpolarized components to be uncorrelated (see \Fig{DuncanPIQU}),
it is natural to expect that $X,Y$=0. However, at the CMB frequencies (where the 
effects of Faraday Rotation \& Depolarization are unimportant) this should 
not be the case.

To study the frequency dependence, we average the 10 multipole bands of the 
Leiden power spectrum measurements together into a single band for each polarization 
type to reduce noise. From these results, we fit the average frequency dependence 
(for the $25\deg$ cut data) as a power law as in \eq{freq_dep} with slope 
$\alpha_E = -1.3$ and $\alpha_B = -1.5$ for $E-$ and $B-$polarization, respectively.


\bigskip
\begin{figure}[tb] 
{\epsfxsize=8.0cm\epsffile{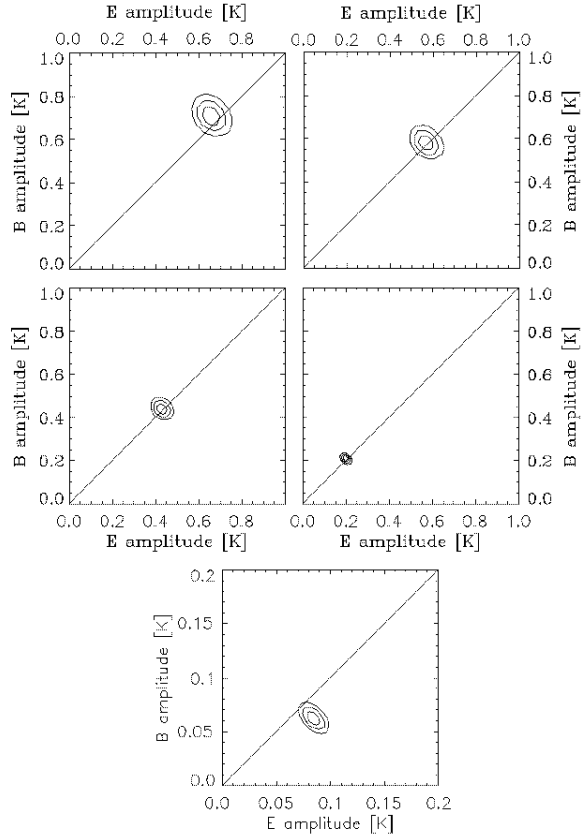}}
{\vskip+0.1cm} 
\caption{\label{likelihood5Leiden}\footnotesize%
	 $E$ and $B$ likelihood values for the Leiden surveys. From top to
	 bottom, and from left to right, the likelihoods are for the 
	 frequencies 408, 465, 610, 820 and 1411~MHz. As the survey's 
	 frequency increases, the Faraday Rotation reduces and we start 
	 to see a slight hint of an $E$-excess. For all likelihoods, the 
	 contours correspond to 68\%, 95\% and 99\% limits. 
	 The diagonal lines correspond to $E=B$.
         }
\end{figure} 

\subsubsection{Is it $E$ or is it $B$?}

An interesting question about polarized foregrounds is how their fluctuations 
separate into $E$ and $B$. Although many authors initially assumed that 
foregrounds would naturally produce equal amounts of $E$ and $B$, Zaldarriaga 
\cite{Zalda01} showed that this need not be the case. There are plausible 
scenarios where the foreground polarization direction could preferentially 
be aligned with or perpendicular to the gradient of polarized intensity,
thereby producing more $E$ than $B$. In contrast, it is more difficult to 
contrive scenarios with more $B$ than $E$, since they require polarizations 
preferentially making a 45$\deg$ angle with the gradient.

Early studies \cite{Baccigalupi00,giardino02} have indicated that $E\approx B$ 
at 2.4~GHz in the Galactic plane. However, these analyses used Fourier transforms 
and spin-2 angular harmonic expansions, respectively, without explicitly 
computing the window functions quantifying the leakage between $E$ and $B$.
This leakage is expected to be important both on the scale of the Parkes 
stripe thickness and on the pixel scale \cite{TC01,Bunn02}, and would have 
the effect of mixing $E$ and $B$ power, reducing any $E/B$ differences 
that may actually be present. Moreover, no study of the $E/B$ ratio has 
ever been done on the large angular scales ($\l \simlt 40$), that are the most 
important for constraining reionization and inflationary gravitational waves.

We therefore perform a likelihood analysis of the Leiden surveys specifically 
focusing on this question, and including an exact treatment of the leakage. 
The likelihood analysis of the data is done with two free parameters 
corresponding to the overall normalization of the $E$ and $B$ power spectra, 
and assuming that they both have the same power law shape given by the slopes 
$\beta_E$ from Table~2.
The results are shown in \fig{likelihood5Leiden}. Note that the $E$ and $B$ 
amplitudes are consistent with being equal to high accuracy at 408, 465, 610 
and 820 MHz. At the highest frequency of 1.4 GHz, however, we see a hint of 
an $E$-excess at the 30\% level, but this is only significant at a level of 
around 95\%.
This hint is intriguing, since it can in principle be given a natural 
physical interpretation. It may be that synchrotron polarization has 
$E>B$ at CMB frequencies, and that Faraday Rotation is hiding this underlying 
asymmetry at low frequencies. If the Faraday effect rotates each polarization 
angle by a for all practical purposes random amount, this will destroy any 
intrinsic alignment between the direction of the polarization and the direction 
of the local intensity gradient and therefore produce equal amounts of $E$ 
and $B$ signal.


\subsubsection{Quantifying the Importance of Faraday Rotation \& Depolarization for the CMB}
\label{quantifying}

The key challenge for modeling synchrotron polarization as a CMB 
foreground is to answer the following question: above which frequency 
are the effects of Faraday Rotation \& Depolarization so small 
that our measurements can be safely extrapolated up to CMB frequencies?
From an analysis of the Leiden surveys, Spoelstra \cite{S84} found 
an upper limit for $RM$ of 35 rad m$^{-2}$. Setting $\Delta\theta=1$ 
rad in \eq{phiequation}, this suggests that the Faraday Rotation 
becomes irrelevant somewhere around 2~GHz. However, considering 
that the determination of $RM$ is poor in many parts of these surveys,
this 2~GHz value is questionable. Moreover, because of the importance of
Depolarization which affects large scales more than small scales, we 
should expect the answer to depend on the angular scale $\l$ considered.

\begin{figure}[tb]
\centerline{\epsfxsize=7.5cm\epsffile{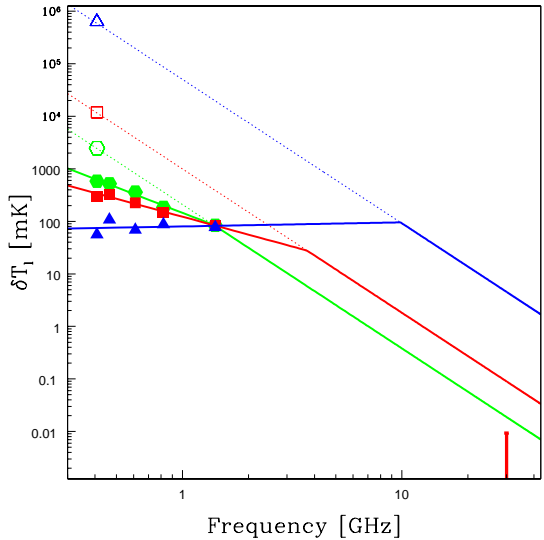}}
{\vskip+0.3cm} 
\caption{\label{nul}\footnotesize%
	 The $E$-polarized (solid) and unpolarized (dashed) power spectra 
	 $\delta T_{\l}$ of Galactic synchrotron emission are plotted as a 
	 function of frequency for multipoles $\l=2$ (blue), $\l=14$ (red) 
	 and $\l=50$ (green) using the fits from Table~2 (corresponding 
	 to a $25\deg$ Galactic cut data). The $T$ curves (dashed) assume 
	 $\alpha=-2.8$.
	 For comparison, the POLAR upper limit of $E<7.4\mu$K 
	 centered in $\l\sim 14$ (see Table 1) is shown in the lower right corner. 
	 Comparing this with the red curve implies either a low synchrotron 
	 polarization percentage or a steeper spectral index (lower $\alpha$).
         }
\end{figure}

Let us now quantify this empirically. \Fig{nul} shows the synchrotron 
power spectra as a function of frequency for a sample of angular scales $\l$.
Using the fits from Table~2 and \Eq{freq_dep} suggests that the 
polarization percentage $p\equiv\delta T_\l^E/\delta T_\l^T$ saturates to 
a constant value for
	$\nu\gg  1$~GHz at $\ell$=50, 
	$\nu\gg  4$~GHz at $\ell$=14 and 
	$\nu\gg 10$~GHz at $\ell$=2. 
This suggest the following universal behavior\footnote{
In the limit of high frequencies (where
Faraday Rotation \& Depolarization vanish), we expect the polarization fraction $p$ to 
become frequency independent. It may still depend on angular scale 
$\l$, however. If it does depend on $\l$, 
there is no fundamental reason why it cannot exceed $100\%$
on some angular scales (even though the polarization at a given {\it point}
is by definition $\le 100\%$) --- imagine, say, a uniform 
synchrotron-emitting plasma with small-scale variations in the magnetic field
direction.
However, \fig{power} and Table 2 show that as the frequency increases, the 
polarized power spectrum gets progressively redder, providing a tantalizing hint
of convergence towards the same power spectrum slope as the total intensity compolent.
It this is actually what happens in the high frequency limit, then the
polarization fraction does indeed become a simple constant. 
}. At high frequencies, where 
the Faraday Rotation \& Depolarization effects are unimportant and the 
polarized fluctuations simply constitute some constant fraction of the 
total fluctuations, we can use the same $\alpha$ for polarized and 
total synchrotron radiation in the CMB range.
However, moving to the left in \fig{nul}, one reaches a critical frequency 
$\nu_*$ below which the Faraday Rotation \& Depolarization effects suppress 
the polarized fluctuations.
At this point, the power law changes asymptotes from a steeper (solid lines) 
to a shallower (dashed lines) power law, and the critical frequency $\nu_*$ 
in which this effect occurs change with the angular scale $\l$. In order words,
whether we can safely extrapolate our results up to CMB frequencies depends 
not only on the frequency but also in the angular scale. For instance, the 
contamination of the CMB quadrupole from Galactic synchrotron polarization can 
only be obtained from extrapolations of data at frequencies exceeding 
$\nu_*\sim 10$~GHz, with $\nu_*$ dropping towards smaller angular 
scales\footnote{
	Due to the fact that we are dealing with cross-correlations, 
	the results presented here should not be biased by systematic errors or
	calibration uncertainties in input data, since they would
	be uncorrelated betweeen the different surveys used.
	Spurious offsets will not cause excess noise either, since 
	we removed the zero-point from each survey before calculating 
  	the cross-correlations.
	}. 

\begin{figure}[tb]
\centerline{\epsfxsize=7.5cm\epsffile{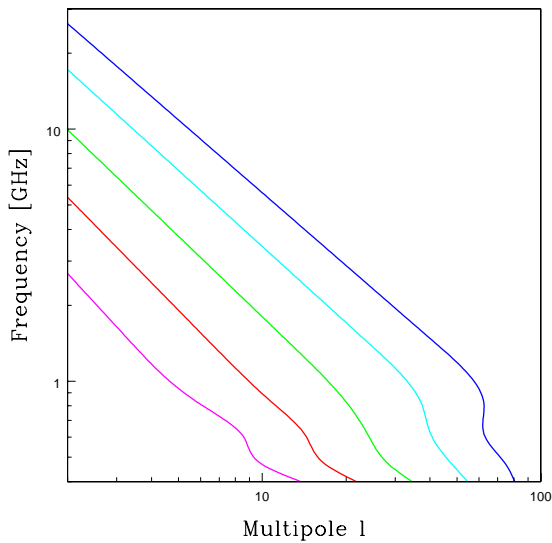}}
{\vskip+0.1cm} 
\caption{\label{saturation}\footnotesize%
         $(\l,\nu)$ plane showing contours of constant 
	 polarization percentage. 
	 From bottom up, the curves are for the 
	 0.01\%, 0.1\%, 1\%, 10\% and 70\%.
        }
\end{figure}

All the information above is summarized in \fig{saturation}, which shows 
contours of constant polarization percentage $p=\delta T_\l^E/\delta T_\l^T$ 
in the two-dimensional $(\l,\nu)$ plane.
In other words, this figure can be interpreted as a contour plot of the
Depolarization. The Depolarization is seen to be negligible at high frequencies 
and on tiny scales, gradually increasing towards the lower left corner 
(towards low frequencies and on large angular scales) where Faraday Rotation 
\& Depolarization effects become important.
 
{\vskip-0.6cm} 
\bigskip

This has important implications. For instance, a nice all-sky simulation of 
synchrotron polarization at CMB frequencies was recently performed assuming 
that the power spectra of $\cos 2\theta$ and $\sin 2\theta$ (where $\theta$ 
is the polarization angle) were frequency independent \cite{giardino02}.
Our results indicate that these two power spectra are dominated by Faraday 
Rotation \& Depolarization effects, which implies that the $E$ and $B$ power 
should be mostly due to changes in polarization angle $\theta$, and not to 
variations in overall intensity -- this precise behavior is also seen by 
\cite{giardino02}.
If Faraday Rotation \& Depolarization effects are indeed dominant, then it 
is not obvious that such frequency extrapolation of the $\cos 2\theta$ and 
$\sin 2\theta$ power spectra are valid. 

\Fig{nul} also shows the POLAR limit of $E<7.4\mu$K from Table 1 (lower 
right corner). Since this limit is centered in $\l\sim 14$, it can be directly 
compared with the middle (red) curve. The noticeable gap between the two 
implies that we get interesting constraints from POLAR on foreground models.
No synchrotron polarization is detected even though the Haslam stripe shows 
substantial synchrotron emission in the POLAR region, so either the synchrotron 
polarization percentage is small or the synchrotron emission falls even more 
steeply towards higher frequencies than the plotted curves indicate.
A spectral index $\alpha=-2.8$ (as shown in the plot) is only allowed if 
the polarization percentage $p$ is lower than 10\%. If $p$=20\%, then 
$\alpha < -3.0$, and almost complete polarization (about 70\% is 
physically possible) would require $\alpha < -3.4$, in poor agreement 
with theoretical and observational indications \cite{banday91,jonas99,roger99}.
In other words, our results suggest a rather low synchrotron polarization 
percentage at CMB frequencies\footnote{
	From the $COBE$/DMR-Haslam cross-correlation results 
	\protect\cite{kogut96a,kogut96b}, we know that the rms Galactic 
	signal of the synchrotron emission is lower than $7.1\mu$K at 53GHz. 
	Note that this value is substantially lower than the one we obtain when
	extrapolating the $\l\sim 14$ curve of \Fig{nul} to the DMR 
	frequencies. This result indicates that $\alpha<-2.8$ or that 
	there is a deviation from the power law behaviour at frequences 
	above a few GHz.
	}.
 

\section{Conclusions}\label{ConclusionsSec}

CMB polarization and its decomposition into $E$ and $B$ modes is a topic
of growing importance and interest in cosmology. In the era of MAP, a key
issue is to estimate the contribution of Galactic foregrounds (more 
specifically, polarized synchrotron emission) to these modes. We have used the 
POLAR experiment and radio surveys in order to quantify this contribution 
at large angular scales.

Using matrix-based quadratic estimator methods, we cross-correlated
POLAR with DMR data and obtained upper limits of 
	$E<7.4\mu K$ and $|X|<11.1\mu K$ 
at 95\% confidence. These upper limits are, unfortunately, to high to place 
intetesting constrains on reionization models. A similar cross-correlation 
analysis was performed by replacing the DMR with the Haslam data, obtaining 
an upper limit of 
	$|X| \simlt 15.4\mu K$ 
at 95\% confidence. 

We also used our quadratic estimator methods to measure the power spectra 
from the Leiden surveys, obtaining the following key results:
\begin{enumerate}
	\item The synchrotron $E$- and $B$-contributions are equal to within 
	      10\% from 408 to 820~MHz, with a hint of $E$-domination at higher 
	      frequencies. One interpretation is that $E>B$ at CMB frequencies
	      but that Faraday Rotation mixes the two at low frequencies.
	\item Faraday Rotation \& Depolarization effects depend not only on 
	      frequency but also on angular scale -- they are important at 
	      low frequencies ($\nu\simlt 10$ GHz) and on large angular scales.
	\item We must take into account Faraday Rotation \& Depolarization 
	      effects when extrapolating radio survey results from low to 
	      high galactic latitudes and from low to high frequencies.
	\item We detect no significant synchrotron $TE$ cross correlation coefficient 
	      ($|r|\simlt 0.2$), but Faraday Rotation could have hidden a substantial 
	      correlation detectable at CMB frequencies.  
	\item Combining the POLAR and radio frequency results, and the fact that 
	      the $E$-polarization of the abundant Haslam signal in the POLAR region 
	      is not detected at 30 GHz, suggests that the synchrotron polarization 
	      percentage at CMB frequencies is rather low.
\end{enumerate}
Experiments such as MAP and Planck will shed significant new light on synchrotron 
polarization and allow better quantification of its impact both on these experiments 
and on ground-based CMB observations.

\bigskip
\bigskip

This work was supported by 
NSF grants AST-0071213 \& AST-0134999 and
NASA grants NAG5-9194 \& NAG5-11099.
MT acknowledges a David and Lucile Packard Foundation fellowship
and a Cottrell Scholarship from Research Corporation.



\end{document}